1

# TO FIND ONE'S WAY OR NOT THROUGH AN UNFAMILIAR ENVIRONMENT[1]

YASMINE BOUMENIR*, FANNY GEORGES*, JEREMIE VALENTIN^, GUY REBILLARD#

& BIRGITTA DRESP-LANGLEY*

*CNRS UMR 5508, Université de Montpellier, CC 048 Place Eugène Bataillon, 34095 Montpellier CEDEX 5, FRANCE

^CNRS FRE 2027 Université de Montpellier, FRANCE

#INSERM U583-IFR76, Université de Montpellier, FRANCE

---

[1] Address correspondence to birgitta.dresp-langley@univ-montp2.fr

*Summary*.— Strategies for finding one's way through an unfamiliar environment may be helped by 2D maps, 3D virtual environments or other navigation aids. Direct experience with the environment prior to navigation is the most natural way of gaining knowledge about routes, directions, and landmarks for the effective construction of spatial representations. This pilot study investigates the relative effectiveness of aids other than direct experience. Experiments were conducted in a large, park-like environment. 24 participants, twelve men and twelve women ranging in age between 22 and 50 years (*M*=32, *SD*=7.4) were divided into three groups of four individuals. Four explored a 2D map of a given route prior to navigation, four were given a silent guided tour by means of an interactive virtual representation, and four acquired direct experience of the real route through a silent guided tour. Participants then had to find the same route again on their own. None of them was familiar with the environment or had visited it before. Itineraries were tracked by means of a GPS device. Twelve observers were given a "simple" route with only one critical turn and the other twelve were given a "complex" route, with six critical turns. Indicators of performance such as the number of wrong turns, times from departure to destination, distances covered and average speed were recorded for each participant through the tracking device. Tracks of three of the authors (26, 28 and 34 years old, two women and one man), who were familiar with the routes, had been recorded a day before to gather reference measures. Results of the naïve participants showed that those who had a direct experience prior to navigation all found their way again on the simple and complex routes. Those who had explored the interactive virtual environment and were given the complex route were unable to find their way again. The findings are interpreted in terms of a problem of relative scale representation in the virtual environment, which conveyed wrong impressions of relative distances between objects along the itinerary, rendering important landmark information useless.

Navigating through novel environments is a complex task one has to accomplish often in daily life. Research has shown that finding one's way through an unfamiliar environment is facilitated by prior experience through direct and guided exposure (e.g. Levine, Warach, & Farah, 1985). However, in real life it is often not possible to have such direct experience with a novel environment before traveling and, therefore, planning which routes to take and which ones to avoid is generally accomplished with the help of indirect information sources and specific navigation aids. Cognitive functions such as attention, perception, and memory are fully solicited for understanding and selecting relevant information that will lead to effective mental representations and spatial understanding of the unknown environment. Spatial understanding may be conveyed through traditional navigation aids such as 2D route maps, or through more sophisticated visualization tools, which more or less faithfully reproduce the real world topology on a computer screen. These different navigation aids may to a greater or lesser extent facilitate finding one's way by providing more or less useful visual interfaces between internal (mental) representations of space and the external environment (Barkowsky & Freksa, 1997). The success of any navigation strategy ultimately depends on how effectively spatial knowledge is made available by a given aid and how this knowledge is then exploited to build working memory representations that enable adequate decision making. These spatial representations have to be reliable enough to allow one to find one's way through a novel environment as quickly as possible without getting lost.

A variety of factors, both personal and environmental, contribute to an individual's ability to find her/his way (Prestopnik & Roskos-Ewoldsen, 2000). Personal factors would include specific characteristics such as sex, relative familiarity with a given environment or with navigating in general, and personal or preferred, strategies. Individuals who preferentially rely on route knowledge, for example, were found to get more easily lost when they happened to deviate from the learned route (Lawton, 1994). While some are good with visual maps, others prefer verbal instructions. Skeletal verbal descriptions of itineraries and landmarks have provided a particularly efficient medium in helping human navigators build cognitive representations of a novel route (Denis, 1997; Denis & Briffault, 1997; Denis, Pazzaglia, Comoldi, & Bertolo, 1999). Environmental factors which may influence an individual's ability to find his/her way would include the relative density of buildings in an area, the availability of meaningful landmarks, and the geometric pattern or layout of streets and their intersections. Noticeable inter-individual differences in exploiting verbally communicated descriptions, compared with visual maps have been reported (Denis, Pazzaglia, Comoldi, & Bertolo, 1999). Being able to effectively use visual maps requires skills to establish object correspondences between the map and the real world topology and to match the map to the environment (Rovine & Weisman, 1989; Newcombe & Huttenlocher, 2000). Such skills may be more or less difficult to acquire. It has been shown, for example, that different individuals tend to encode spatial knowledge from visual maps in different ways, which was made evident when map-based strategies were compared with strategies based on direct experience through guided environmental exposure (Levine, Warach, & Farah, 1985). Other studies have shown that subjects may benefit from 2D visual maps equally well as from direct experience (e.g. Ishikawaa, Fujiwarab, Osamu, Imaic, & Okaboc, 2008). Such benefit was reflected through shorter distances covered, shorter times taken to get to the target location and a lesser number of stops during navigation compared with a control situation, where subjects had to navigate without any aid or prior experience.

      Three types of knowledge are relevant to route planning strategies: landmark knowledge, route knowledge, and survey knowledge. The spatial understanding derived from landmarks,

routes, and their configurations is critical to elaborate effective criteria for deciding where to go and when (Siegel & White, 1975). To which extent each type of such spatial knowledge enables individuals to build reliable internal representations of external space remains to be clarified (Xiaolong, 2008). Little is still known about the relative advantage of route, landmark, or survey knowledge in different environments such as forests, cities, or parks. The relative complexity of itineraries should also have a determinant influence, yet, not much is known about how route complexity would affect the construction of different types of spatial knowledge.

      Route knowledge generally relates to the geometry and spatial layout of an itinerary. It allows one to decide where to go next, where to turn, or how to get from one location to another. It has been reported (e.g. Sun, Chan, & Campos, 2004) that information about spatial layout of routes and intersections communicated through 2D visual maps can be more efficient for subsequent navigation than prior exposure or direct experience, or a virtual representation of the real environment. The observation of a specific alignment effect (Sun, Chan, & Campos, 2004) with 2D visual maps suggests that the precision with which the geometry of routes and their spatial configuration are represented on the map is of critical importance.

      Landmark knowledge relates to representations of distinct features of prominent objects such as buildings, monuments, or shops encountered along a given route. Such representations require place knowledge ("where am I") to be exploited for constructing routes and for navigating ("where do I go next"). Some have used the term landmark in a very general way to refer to any decision point in space (e.g. Golledge, 1999). Landmarks may help organize space in terms of reference points, or choice points, where navigational decisions are made, and to identify origin and destination. They can represent clusters of objects at a higher level of abstraction or scale, or serve as anchors for understanding local spatial relations. Landmarks may

enable the encoding of spatial relations between objects and routes, enhance prior route knowledge, and facilitate the construction of a cognitive map. Exploiting landmarks to build survey knowledge of an environment or itinerary may facilitate spatial orientation, the discovery of new or alternative routes, and the discrimination between different areas or regions (e.g. Golledge, 1999; Janzen, Janse, & Turrenout, 2007). Mental representations of routes, landmarks, and regions constitute what Tolman (1948) referred to as cognitive maps of the physical environment. Cognitive maps appear to be constructed hierarchically, grouping different features of space and complex spatial layouts into different levels of representation, thereby generating an understanding of surrounding space.

    Spatial understanding derived from virtual representations of a real world environment appears to largely rely on the visual representation of key landmarks (e.g. Ohmi, 1996), which users may exploit to scale the virtual space. An important drawback of virtual representations of a real world environment is related to the fact that such representations restrict the perceptual field to the size of the screen and deprive the user of vestibular, kinesthetic and proprioceptive cues, which are highly useful for navigating in the real world (Ohmi, 1996; Berthoz & Viaud-Delmon, 1999). However, some studies have shown that spatial knowledge acquired through learning based on virtual representations transfers quite well to subsequent navigation in the real world (Arthur, Hancock, & Chrysler, 1997; Witmer, Bailey, Knerr, & Parsons, 1996). Others (Liben, 2001; Meilinger, Knauff, & Bulthoff, 2008) have shown that 2D visual maps and virtual representations on a screen can be equally useful for constructing reliable spatial knowledge. More recently, Ishikawa, Fujiwarab, Osamu-Imaic, & Okaboc (2008) have found that GPS-based mobile systems are ineffective in helping observers find their way in comparison with prior direct guided experience, or 2D visual map exploration prior to navigation.

This study investigated the relative effectiveness of route, landmark, and survey knowledge on the success of spontaneous human navigation in a park-like environment with which none of the observers was familiar. The three different types of spatial knowledge were selectively made available to participants prior to navigation through one of three different aids. One group was given a direct guided visit along the real world route before they had to navigate on their own. In this potentially optimal familiarization condition, route, landmark, and survey knowledge was made available. A second group was given a simple 2D visual map, showing the spatial layout of routes and their intersections and allowing participants of this group to build a 2D geometric representation of their itinerary without any information about landmarks. It was made sure that relative distances between routes and intersections were adequately scaled on the maps. Another group of participants was given an interactive virtual tour of their route prior to navigation in the real environment. The virtual tour consisted of step-by-step navigation through a sequence of panoramic, photorealistic representations of the itinerary. It gave participants of this group access to route knowledge and potentially significant landmark knowledge. After a given familiarization procedure, observers had to find their way once again and on their own in the real world environment. The effect of the spatial complexity of the route that had to be found again was investigated by confronting half of the observers of each group with a relatively simple, the other half with a more complex itinerary.

## Methods

Different groups of participants were requested to navigate along two different, unfamiliar routes in a park-like environment forming a large cemetery (Père Lachaise in Paris), with a complex network of smaller and larger human-made paths for pedestrians. Figure 1 shows a 2D representation of the location and study routes, with photographs of potential landmarks along these routes. One of the routes we chose may be described as "simple" since it covered a relatively short distance, represented by a straightforward linear geometry with a single sharp left-turn into a wide and relatively open route with only one more intersection. The other route may be described as "complex", covering a distance twice as long, represented by a geometry which was linear first and then, after a sharp right turn, led into a dense forest with a network of many smaller routes and paths with multiple intersections. Individuals had to navigate from memory to get from the point of departure to one of the two target locations after having studied a 2D map, having been given a silent guided tour to the location (direct experience) or having been given the opportunity of virtual (*Quick Time*) mouse navigation along the given route on a computer screen through the web application www.perelachaise.com.

*Participants*

24 adults, 12 men and 12 women, all volunteers, participated in this study. They were selected from a population of healthy, fully mobile and professionally active individuals on the basis of an advertisement posted to acquaintances and their families. None of them reported any neurological or motor deficit. Their ages ranged from 22 to 50 (*M*=32, *SD*=7.4). None of the participants was familiar with the study location. To collect reference measures, three of the authors participated as so-called "experts". They had studied and learned the itineraries described here, and were able to navigate along the simple and complex routes swiftly without getting lost.

*Measures*

*Study area and routes.*— A study area was chosen in the Père Lachaise cemetery in Paris. Two different routes were selected, with a common starting point at the main entrance. The simple route required a single left-turn to get from the point of departure to the target location (a fountain) through a relatively open space with few routes and intersections. The complex route was twice as long and required several choices as to where to go next to reach the target location (another fountain) through a complex network of small paths with intersections and a dense forest. An intermediate key location (a chapel) on that route was mentioned to the participants. The target locations (fountains) were not visible from the starting points of either route. The intermediate key location was not visible from the starting point of the complex route. In the simple route conditions, participants were requested to find their way to the fountain at the target location. In the complex route conditions, they were requested to find their way to the fountain at the target location by passing by a chapel at an intermediate location.

*GPS tracking.* — For tracking individual paths, mobile devices equipped with internal GPS receivers were used. *GPSed* software permitted exporting the data recorded during navigation into the generic "GPX" format for statistical analyses. GPS tracking allowed recordings of one track per second, generating reliable data relative to distances covered by an observer and his/her average speed. The trees in the study area were not an obstacle to GPS tracking, the density of their foliage being at its lowest at the time (November 2008).

*Silent guided tours.* — Eight of the 24 observers were given a silent guided tour of the route they had to travel again afterwards on their own. Four observers (two men, two women) were guided to the fountain along the simple route, four others (two women, two men) to the other fountain along the complex route. The silent guided visit consisted of accompanying the individual from the point of departure to the target location of a given path and back. Sculptures, tombstones and other potential landmarks such as panels indicating the divisions of the cemetery, or the names of paths, are encountered on both the simple and the complex route. At the starting point of the routes, the main entrance, a gate opens on a large road leading to an imposing war memorial situated along the track leading to the target location of the complex route. A key location on this complex route was a large chapel, with a square with flowerbeds and seats in front of it. Participants traveling the complex route were instructed to "go to the fountain by passing in front of the chapel". It may be worth mentioning that, at the time of this study, most of the tombs were decorated with particularly distinctive ornaments, such as wreaths with ribbons featuring names and other text. These were not permanent features and therefore not present in the virtual guided tour through the cemetery.

*Virtual guided tours*. — Another group of eight participants was given a virtual guided tour by one of the authors prior to navigation. The tour consisted of silently guided mouse navigation through virtual representations on a computer screen of the routes they had to find afterwards on their own in the real environment. Four observers (two men, two women) were guided along the simple route from departure to target location and back, four others (two women, two men) were guided along the complex route and back. In this virtual guided tour condition, a given route was represented interactively through successive 360° panoramic views (*Quick-Time* technology) flashed on a computer screen. Successive mouse clicks on arrows indicating the directions available allowed step-by-step navigation through the successive 3-D views. Guidance merely consisted of pointing out to the participant where to click next. The original web interface features a tabular map of the cemetery, next to the frame provided for virtual viewing. In the present study, the full screen zoom (on a 15' monitor) was activated to hide this map from view. Interactive features of the virtual environment allowed participants to generate 360° panoramic views of the surroundings of a given route. This could be achieved by moving the mouse to the left or to the right at any given position before continuing on the route by clicking on red arrows indicating directions. Arrows did not indicate turns, they only pointed straight ahead, which makes it particularly difficult for anyone not familiar with the real world route to scale the environment and to generate reliable estimates of relative distances between potential landmarks. Also, during the virtual guided tours, the viewpoint of a given panoramic image sometimes had to be re-adjusted by the tour guide to make a new direction available to a participant for further navigation.

*Two dimensional (2D) route maps*. — Eight (four males, two females) of the 24 subjects were given a 2D map of either the simple or the complex route before they had to travel on it to find a given fountain. Maps were scaled appropriately, reproducing relative distances between locations reliably, and printed out on A4 sheets. Starting points, intermediate locations, and target locations were indicated on these maps. The global network of routes was shown, but trajectories to follow were not highlighted. The maps were handed out to participants for exploration prior to navigation and were then removed after two minutes. This time was deemed largely sufficient by all of the participants for understanding and memorizing the itineraries.

*Procedures*

After a direct experience, a virtual tour, or a two minute exploration of a given 2D route map, participants of the different groups were taken individually to the starting point of the simple and complex routes at the main entrance. A small bag containing a GPS tracking device was placed around their necks, and participants were instructed not to touch the device. All of them had their own mobile phones to call one of the investigators in case they were not able to find their way to a target location and did not manage to find the way back to the main entrance. Individuals who found their way to their target location were met by an investigator, who collected their GPS device. Participants who got lost were searched for by another investigator, and accompanied back to the entrance. At the end of the experiment, participants were interviewed about their individual strategies and about specific landmarks that may have seemed useful to them.

*Data analysis*

The performance of the participants in the different conditions was assessed by exploiting individual GPS data relative to distances covered, times taken from departure to destination, and the number of navigation errors in terms of wrong turns taken. The same number of male and female subjects was tested in a given experimental condition and performances were analyzed as a function of the type of navigation aid used prior to navigation, and the complexity of the route traveled. The individual data recorded previously from the three authors, who were well familiar with the itineraries, were analyzed for comparison.

Results

*Visualization of individual GPS tracks*.— All participants reached their target location and no noticeable differences in performances were observed in the simple route conditions. This becomes evident when comparing the individual GPS tracks visualized for the simple routes, and plotted together with the GPS track recorded for one of the authors. The complex route, on the other hand, produced large deviations of the participants' tracks from the optimal route, which is made evident by comparing the individual GPS tracks of the different participants visualized for the complex route conditions. Individual GPS tracks are shown in Figure 2. Further analyses of performance in terms of complementary parameters, such as times from departure, distances covered, average speed, and number of wrong turns, were then performed as a function of the type of navigation aid used prior to navigation.

*Time, distance, average speed and number of wrong turns*. — Distances covered, navigation speed, and other parameters were analysed using *EveryTrial* software, which allows processing the GPX data explored through *GPSed*. For each participant, the time taken from departure, total distance traveled, and average walking speed was computed for a given itinerary. The number of times a participant made a wrong turn leading him/her away from the path towards the destination (turning left or right instead of continuing straight ahead, for example) was calculated. Table 1 shows these data as a function of sex, type of aid used prior to navigation, and type of route traveled. Table 2 shows means, standard deviations and 95% confidence intervals as a function of the type of navigation aid used prior to navigation and the type of route travelled.

For participants who had to travel the simple route, no systematic differences in study parameters between men and women or different types of navigation aid were observed (see Table 1). Success rates were 100% in all cases, indicating that all found their way again on the simple route (see Table 1). Times from departure to destination on the simple route varied between 150 and 333 seconds, distances traveled between 225 and 343 meters, the number of wrong turns between 0 and 1, and the average travel speed between 2.7 and 6.2 kilometers per hour. Means, standard deviations and 95% confidence intervals are shown in the upper part of Table 2. The data recorded for participants traveling the simple route were not noticeably different from the data of the three authors, shown in Table 3 for comparison, who were well familiar with the route.

For participants who had to travel the complex route, remarkably longer times from departure to destination and longer distances covered were recorded for women in the 2D map condition (see Table 1). In the other conditions with direct experience and virtual guided tour prior navigation, the times and distances were longer for men (see Table 1). Significant differences in study parameters were observed between conditions with different types of navigation aid. The best performances, with the smallest number of wrong turns, the shortest times from departure to destination, shortest distances traveled, and fastest average speed on the complex route were observed after direct experience with the real itinerary (see Tables 1 and 2). The success rate was 100% in this condition, indicating that all participants found their way again after a direct guided visit prior to navigation. Performances closely approach those of the three authors who were familiar with the route (see Table 3), which were, as would be expected, the best recorded on the complex route.

    The worst performances, with the highest number of wrong turns, the longest times from departure, the longest distances traveled and a success rate of zero were observed after a virtual tour prior to navigation on the complex route (see Tables 1 and 2). All participants in this condition got lost and did not find the target location again, which readily explains the unusually long times from departure and distances traveled recorded in this condition (see Tables 1 and 2). The 2D map condition produced recordings of parameters indicating performance levels above those of the virtual tour condition and below those of the direct experience condition (see Tables 1 and 2), with a 50 % succes rate indicating that half of the participants in this condition did not find the target location again after having explored a 2D map.

*Analysis of variance*. — Further statistical analyses (ANOVA and post-hoc t-tests) were performed, comparing means of performance parameters recorded in the different conditions for participants who had to travel the complex route (as shown in Table 2). The results of these further analyses are summarized in Table 4. 2x3 ANOVA comparing between means (N=6) from the three navigation aid conditions across the two route conditions indicated a statistically significant effect of the type of navigation aid on the number of wrong turns between study populations (see Table 4 for F-values and probability boundaries). 1x3 ANOVA comparing means (N=3) from the independent study groups within the complex route condition gave statistically significant effects of the type of navigation aid on the number of wrong turns, times from departure (effect at significance margin) and average speed (see Table 4 for F-values and probability boundaries). One-to-one post-hoc comparisons (t-tests) between individual data of the study groups (N=4 per group) from the complex route condition gave statistically significant effects on the number of wrong turns, the times from departure and the average travel speed comparing between the direct experience and the virtual guided tour groups (see Table 4 for t-values and probability boundaries). Comparisons between the direct experience and 2D plan groups of the complex route condition indicated a statistically significant difference in average travel speed.

*Reported preferences and strategies*. — Individual replies to the questionnaires that were filled out by the participants during the interviews conducted immediately after testing, gave some insight into the strategies observers seemed to have been employing while navigating. Participants from the groups with direct experience or a guided virtual tour prior to navigation reported having memorized specific landmarks, such as gravestones, signposts, staircases, or chapels, for finding their way again afterwards, especially on the complex route. Participants from the 2D map groups reported having memorized orientations and their succession, such as left, right, and straight ahead, for subsequent navigation.

## Discussion

The present field study addresses ecologically important aspects of human behavior by clarifying some of the factors which account for the success or failure of spontaneous human navigation in unfamiliar real world conditions. In everyday life, the success of such navigation often depends on indirect information sources or navigation aids, which may not always be effective in communicating the spatial knowledge that is relevant for subsequent navigation. Here, the relative effectiveness of 2D maps, a guided direct visit, or the guided use of a Quick Time virtual reality tool, used prior to navigation from a departure point to a specific target location, was investigated. The relative complexity of the route that had to be traveled was also varied. Results of the study confirm the superiority of direct experience with, or guided exposure to, the real environment as an aid for subsequent navigation, on the simple as well as on the substantially more complex itineraries investigated here. Moreover, the findings clarify that the photorealistic character of 3D virtual environments, reproducing the real world itinerary in an apparently truthful manner on a computer screen, does not *per se* provide spatial knowledge for a successful navigation in the real world environment. Other factors related to scale, direction, and relative distances between routes and objects encountered along them, appear more relevant. This may explain why simple 2D maps were found significantly more efficient than the seemingly sophisticated 3D representations. In a single view, 2D maps provide essential geometric information about routes, configurations, directions, and relative distances and although they are not photorealistic reproductions of the real environment, they convey the visual spatial information that is necessary to elaborate skeletal representations of space. Such representations are linked to perceptual and mnemonic processes relevant to what has been termed cognitive mapping (e.g. Golledge, 1999).

Landmarks carry symbolic information that is important for cognitive mapping because symbols may trigger useful associative processes in long-term memory. Evidence from functional magnetic resonance imaging (fMRI) studies, for example, has shown that the structure in the human brain (the parahippocampal gyrus) that ensures the long-term encoding of the spatial relevance of landmarks is activated differentially in competent navigators and poor ones (Janzen, Jansen, & Turennout, 2007). This result tends to suggest that landmark learning is a critical aspect of becoming a good navigator. It could be particularly important when the geometry or spatial layout of a novel environment is too complex to be exploited effectively for navigating from memory (cf. the complex route conditions in this study).

Direct experience with a novel environment prior to navigation gives ready access to knowledge about landmarks, and so should, in principle, prior experience with the equivalent virtual environment. However, panoramic step-by-step viewing of sequences of virtual representations in our study, although it provided participants with the same sequence of landmarks as the direct experience, was quite clearly of no help. This is likely to be due to the fact that participants were unable to effectively use the landmark information provided, or to implement this information into a valid spatial representation of the complex route. This could be explained by the difference in scale between the virtually generated views and the real world itinerary, and by the fact that the panoramic step-by-step viewing of the route and landmarks along it made it difficult, or impossible, to reliably assess relative distances between critical turns and symbols. Another potential disadvantage of the virtual navigation tool, especially in the complex route condition, may relate to the fact that directions were given by arrows pointing always straight ahead on the screen. This implies that participants had to rely on linear representations of space when navigating from memory while the real environment of the complex route followed a complex double-curved geometry.

The interpretation of the findings here sheds new light on current knowledge about navigation strategies, identified in previous studies (e.g. O'Keefe & Nadel, 1978; Russell & Ward, 1982; Passini, 1984; Lawton, 1994; Schlender, Peters, & Wienhöfer, 2000). One of the most common strategies for navigating from memory has been described in terms of route sequencing operations, where instructions one-by-one enable an individual to get from a given place to another until a target is reached. This type of strategy exploits landmarks with an essentially local focus, typified by directions to turn right, left, or continue straight ahead at particular landmarks. Being able to assess how far away a given landmark is from another may be critically important here and, if such is the case, any virtual environment that gives wrong or imprecise impressions of relative distances between salient landmarks is doomed to fail as an aid for navigating from memory. Participants in our study also had to navigate from memory and it can be assumed, especially in the light of their answers to the questionnaire, that they had recourse to route sequencing strategies using critical landmarks in all but the 2D map conditions.

Whether there are significant differences between men and women in solving spatial problems has been subject to debate for some time (e.g. Allen & Hogeland, 1978; Prestopnik & Roskos-Ewoldsen, 2000). Sex differences represent an interesting issue that could not be addressed in this pilot study. It was ensured that the number of men and women participants was balanced within and across conditions. This was preferable given the relatively small number of volunteers found and, although some of the results reported here may suggest sex differences for certain parameters, by showing, for example, that the two women who explored 2D maps prior to

19navigation on the complex route took considerably more time and covered noticeably larger distances than the two men in the same condition, statistical conclusions could not be drawn. The question of possible differences between men and women with regard to spatial abilities is highly controversial (e.g. Voyer, Voyer & Bryden, 1995). Sex differences have not been reported systematically in studies using spatial tasks but some authors have found superior performances for men (e.g. Allen & Hogeland, 1978; McGee, 1979; Linn & Petersen, 1985). Some of the factors which may contribute to explaining such results have been addressed in studies by Lawton (1994, 1996) and Prestopnik & Roskos-Ewoldsen (2000), among others. It has been argued that at least part of the sex differences favouring the performance of men in spatial tasks could be related to differences in preferred navigation strategies. Men, for example, were reported to prefer survey strategies and to rely on cardinal directions (North, South, East, and West) for navigating, whereas women appeared to prefer relying on representations of the spatial layout of itineraries and their local and global configuration.

20

22

**FIGURE AND TABLE CAPTIONS**

FIGURE 1:

Representation of simple and complex route geometries on a 2D colored map (different from the monochrome maps used in one of the conditions in this study) of the study location, with photographs of potential landmarks along the two itineraries.

FIGURE 2:

Individual GPS tracks of the paths traveled by participants of the different study groups. Tracks in the simple route condition (upper panel) and the complex route condition (lower panel) are printed as a function of the type of aid used prior to navigation. Images on the left show tracks of two men (H1, H2) and two women (F1, F2), who had a direct experience prior to navigation. Images in the middle show tracks of participants who were given 2D maps prior navigation. Images on the right show tracks of participants who had been given a guided virtual tour. Tracks of individuals ("experts") who were well familiar with the routes are plotted in red for comparison with the tracks of the naïve participants. The scales of images showing simple and complex routes are not identical, and the representations here only serve for a global comparison between the tracks of participants traveling the same route.

TABLE 1:

Parameters of navigation performance on the simple and complex routes are given as a function of sex and type of navigation aid used prior to navigation.

TABLE 2:

Means, standard deviations, and 95% confidence intervals for the different parameters as a function of type of route traveled and type of navigation aid used prior to navigation.

TABLE 3:

Means, standard deviations, and 95% confidence intervals for parameters recorded for the individuals ("experts") who were familiar with the routes.

TABLE 4:

F-values and probability boundaries are given for 2x3 and 1x3 comparisons between means for number of wrong turns, time from departure and average travel speed in the complex route condition. The lower half of the table gives t-values and probability boundaries for 1x1 post-hoc comparisons between groups within the complex route condition.

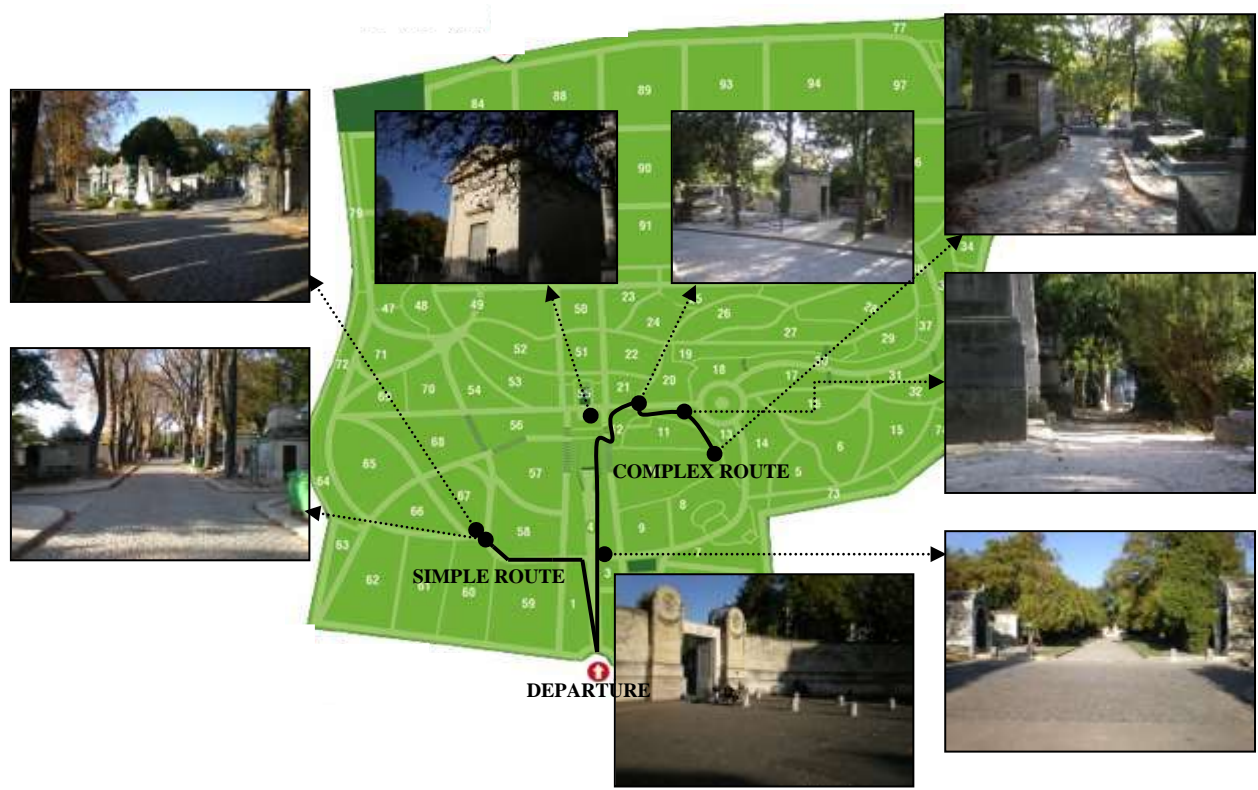

FIGURE 1

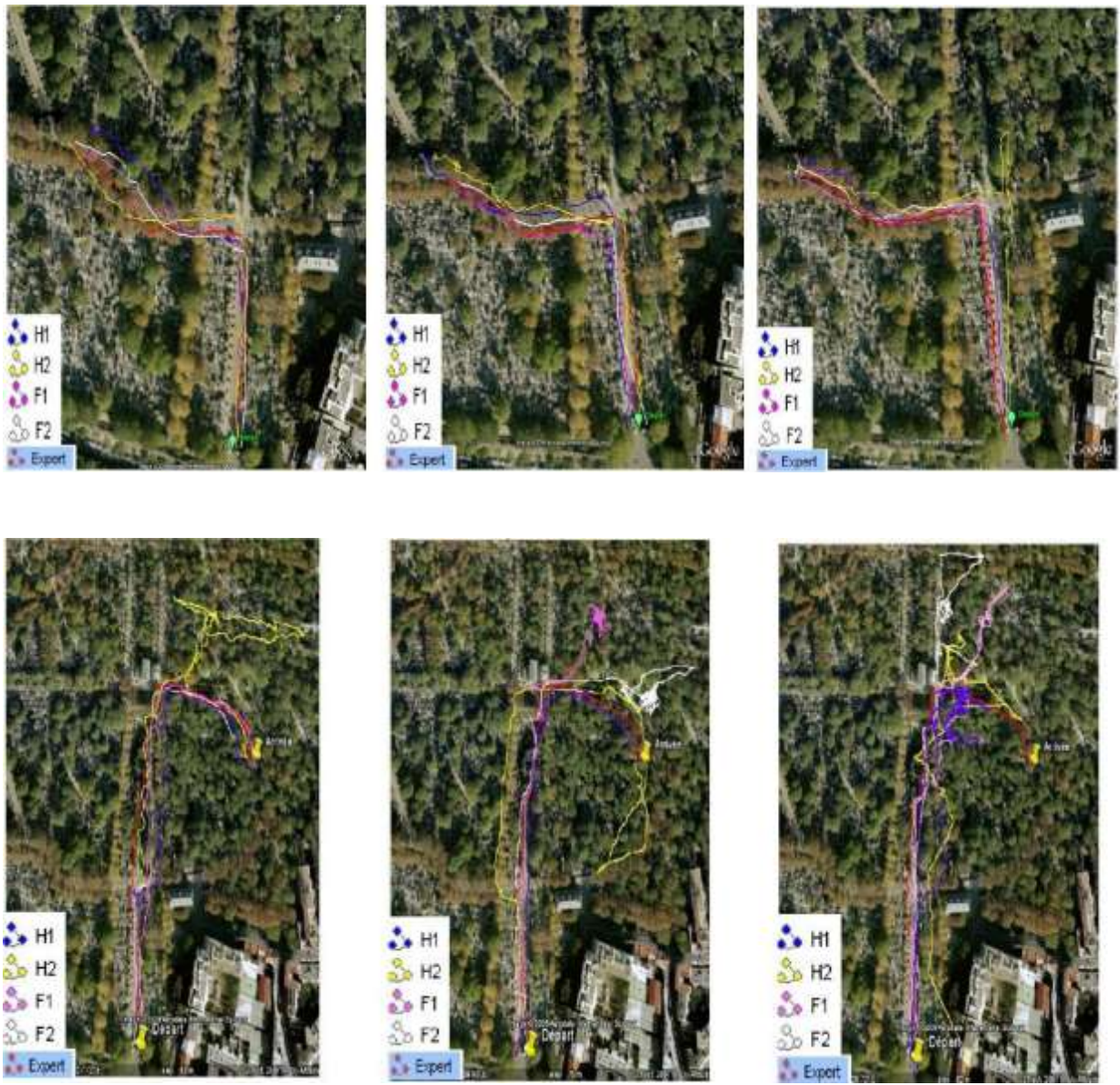

FIGURE 2

|  | 2D plan | | Direct experience | | Virtual tour | |
| --- | --- | --- | --- | --- | --- | --- |
| **Participants on simple route** | ♂ N=2 Age: 35+34 | ♀ N=2 Age: 30+25 | ♂ N=2 Age: 26+29 | ♀ N=2 Age: 32+50 | ♂ N=2 Age: 48+28 | ♀ N=2 Age: 33+22 |
| wrong turns | 0 | 0.5 | 0 | 0 | 0 | 0.5 |
| time from departure (s) | 210 | 251 | 192 | 195 | 280 | 188 |
| speed (km/h) | 5.9 | 3.6 | 4.8 | 3.5 | 3.9 | 5.3 |
| distance travelled (m) | 344 | 236 | 246 | 247 | 293 | 275 |
| success rate | 100% | 100% | 100% | 100% | 100% | 100% |
| **Participants on complex route** | ♂ N=2 Age: 29+24 | ♀ N=2 Age: 34+30 | ♂ N=2 Age: 37+45 | ♀ N=2 Age: 38+24 | ♂ N=2 Age: 29+23 | ♀ N=2 Age: 29+33 |
| wrong turns | 1 | 9 | 2 | 0 | 10 | 10 |
| time from departure (s) | 573 | 1288 | 637 | 351 | 1276 | 1006 |
| speed (km/h) | 3.9 | 3.1 | 4.9 | 4.6 | 3.3 | 4.0 |
| distance travelled (m) | 634 | 1121 | 900 | 445 | 1207 | 1050 |
| success rate | 50% | 50% | 100% | 100% | 0% | 0% |

TABLE 1

|  | 2D plan | | Direct experience | | Virtual tour | |
|---|---|---|---|---|---|---|
| **Simple route** | **M/SD** | **95% conf.** | **M/SD** | **95% conf.** | **M/SD** | **95% conf.** |
| wrong turns | 0.25/0.5 | 0.79 | 0/0 | 0 | 0.25/0.5 | 0.79 |
| time from departure (s) | 230.3/60.4 | 96.2 | 193.7/35.9 | 57.1 | 233.8/69.3 | 110 |
| speed (km/h) | 4.7/1.52 | 2.42 | 4.2/1.17 | 1.86 | 4.6/0.92 | 1.46 |
| distance (m) | 286.6/66.2 | 105 | 246.3/9.3 | 14.8 | 285/26.3 | 41.8 |
| **Complex route** | **M/SD** | **95% conf.** | **M/SD** | **95% conf.** | **M/SD** | **95% conf.** |
| wrong turns | 5/2.5 | 9.8 | 1.25/2.5 | 3.9 | 10/4.4 | 6.9 |
| time from departure (s) | 930/366 | 642 | 493.7/262 | 416 | 1141/170 | 270 |
| speed (km/h) | 3.5/0.50 | 0.80 | 4.8/0.62 | 0.99 | 3.7/0.66 | 1.06 |
| distance (m) | 877/344 | 547 | 673/420 | 668 | 1129/259 | 412 |

TABLE 2

|  | Experts (N=3) | |
|---|---|---|
| **Simple route** | M/SD | 95% confidence |
| wrong turns | 0/0 | 0 |
| time from departure (s) | 172/34.7 | 86 |
| speed (km/h) | 5.8/0.66 | 1.65 |
| distance (m) | 254/12.2 | 30.2 |
| **Complex route** | M/SD | 95% confidence |
| wrong turns | 0/0 | 0 |
| time from departure (s) | 379/94.6 | 234 |
| speed (km/h) | 4.6/1.11 | 2.76 |
| distance (m) | 461/18.2 | 45.1 |

TABLE 3

| 1. ANOVA (F-statistics) | Number of wrong turns | Time from departure (*s*) | Average speed (*km/h*) |
|---|---|---|---|
| 2x3 comparison between means (N=6) for the two route conditions and three types of navigation aid | F(1,5)=11.86, **p=.0002;** | **n.a.** | F(1,5)=1.28, p=.3144 NS; |
| 1x3 comparison between means (N=3) for independent groups within **simple route** condition | F(1,2)=0.49, p=.6247 NS; | F(1,2)=0.61, p=.5665 NS; | F(1,2)=0.19, p=.8337 NS; |
| 1x3 comparison between means (N=3) for independent groups within **complex** route condition | F(1,2)=7.26, **p=.0133;** | F(1,2)=4.16, **p=.0527;** significance margin | F(1,2)=5.495, **p=.0276;** |
| **2. Post-hoc analyses (t-tests)** 1x1 comparisons between individual data of the three independent groups (N=4 per group) within the **complex route** condition: direct experience *vs* 2D plan | t(1,6)=1.63, p=.1548 NS; | t(1,6)=1.90, p=.1055 NS; | t(1,6)=3.08, **p=.0217;** |
| direct experience *vs* virtual tour | t(1,6)=3.797, **p=.0090;** | t(1,6)=2.83, **p=.0301;** | t(1,6)=2.61, **p=.0404**; |
| 2D plan *vs* virtual tour | t(1,6)=2.17, p=.0731 NS; | t(1,6)=0.92, p=.3922 NS; | t(1,6)=0.47, p=.6525 NS; |

TABLE 4